\documentclass[aps,prb,preprint,superscriptaddress,floatfix,showpacs,showkeys,amsmath,amssymb]{revtex4-1}

\usepackage[dvips]{graphicx}

\usepackage[colorlinks=true,linkcolor=blue,citecolor=blue]{hyperref}

\begin{document}


\title{Andreev experiments on superconductor/ferromagnet point contacts}


\author{S. Bouvron}
\altaffiliation{now at: Fachbereich Physik, Universit\"at Konstanz, 78457 Konstanz, Germany }
\author{M. Stokmaier}
\altaffiliation{now at: Institute for Nuclear and Energy Technologies (IKET), Karlsruhe Institute of Technology (KIT), 76131 Karlsruhe, Germany}
\author{M. Marz}
\affiliation{Physikalisches Institut, Karlsruher Institut f\"ur Technologie (KIT), 76131 Karlsruhe, Germany}
\author{G. Goll}
\email{gernot.goll@kit.edu}
\affiliation{Physikalisches Institut, Karlsruher Institut f\"ur Technologie (KIT), 76131 Karlsruhe, Germany}
\affiliation{DFG-Center for Functional Nanostructures, Karlsruher Institut f\"ur Technologie (KIT), 76131 Karlsruhe, Germany}

\date{December 18, 2012}

\begin{abstract}
Andreev reflection is a smart tool to investigate the spin polarisation $P$ of the current through point contacts between a superconductor and a ferromagnet. We compare different models to extract $P$ from experimental data and investigate the dependence of $P$ on different contact parameters.
\end{abstract}


\maketitle

Following the pioneering work of Igor Yanson and his group at the Institute for Low Temperature Physics and Engineering of the Ukrainian Academy of Sciences (ILTPE NASU) on tiny metallic contacts between two metal electrodes,~\cite{Yanson:1974} point-contact spectroscopy (PCS) has become a powerful method to study the interactions of ballistic electrons with other excitations in metals.~\cite{NaidyukPCS:2005} The interpretation of the observed characteristics in point-contact (PC) spectra is usually difficult. These difficulties are frequently inherent in the fabrication of point contacts. In many cases contacts are made by the needle-anvil or shear technique in which two sharpened metal pieces are brought into a gentle touch until a conductive contact is formed. Those contacts are  microscopically not well-defined with respect to contact size, geometry, and structure of the metallic nanobridge, and with respect to the local electronic parameters such as the mean free path in the immediate contact region. The only control parameter is the contact resistance, and hence, it is challenging to identify the relevant transport regime free of doubt. Usually Sharvin's \cite{Sharvin:1965} or Wexler's \cite{Wexler:1966} formulae for the ballistic and diffusive transport regime, respectively, are used to infer a PC size estimate from the measured PC resistance. Only recently,~\cite{gra12} direct scanning electron microscopy (SEM) measurements of the nanocontact size of nanostructured point contacts allowed for the first time a direct comparison with theoretical models for contact-size estimates of heterocontacts. The semiclassical models yield reasonable values for the PC radius $a$ as long as the correct transport regime is determined by taking into account the local transport parameters of the individual contact. Of course, this requires a careful characterisation of the samples with respect to the local resistivity and the local mean free path.

Among the rich variety of solid-state problems investigated by point-contact spectroscopy the study of superconductor-metal contacts contributes a significant portion. Nowadays point-contact spectroscopy is an important tool to explore the symmetry and nodal structure of the energy gap $\Delta$ of conventional and unconventional superconductors.~\cite{gol05} When the temperature is lowered below the superconducting transition temperature $T_c$ of the superconducting electrode of a superconductor (S)/ normal metal (N) point contact Andreev reflection~\cite{and64} of charge carriers at the S/N interface occurs. Andreev reflection  leads to minima at $V\approx \pm \Delta /e$ in the differential resistance $dV/dI$ as a function of applied bias $V$, i.\,e. maxima in the corresponding conductance curves $G(V)=(dI/dV)(V)$, and thus allows determination of the gap size, also while varying temperature and magnetic field, respectively.~\cite{zai80,blo82} 

A new pitch came into the field when Andreev reflection was used to extract the spin polarisation $P$ of the current through superconductor/ferromagnet (F) point contacts.~\cite{Soulen:1998,upa98} Knowledge of the spin polarisation of possible materials for spin-electronic devices is a key issue for spintronics.~\cite{pri95} An efficient spin injection is of central importance for utilizing the spin degree of freedom as a new functionality in spin-electronic devices. The spin polarisation $P$ of ferromagnets can be measured by various techniques including photoemission,~\cite{bus71} spin-dependent tunnelling,~\cite{mes94} and point-contact Andreev reflection (PCAR).~\cite{jon95} For a quantitative analysis of the results, however, one has to be aware of the different nature of the quantities measured by each technique. The spin polarisation, defined by the difference of spin-up and spin-down density-of-states, is typically measured with spin-polarised photoemission while the spin polarisation of the transport current is obtained, e.\,g. in PCAR experiments,~\cite{maz99} which in turn is distinctly different from the spin polarisation of the density-of-states resulting from tunnelling experiments.~\cite{mes94} An issue of considerable importance is how the spin polarisation obtained by Andreev reflection is related to the ferromagnet's bulk spin polarisation.~\cite{xia02}

Already a variety of materials have been investigated including the ferromagnetic elements Fe, Co, and Ni and several alloys mainly with Al, Nb, or Pb as a superconducting counter electrode.~\cite{Soulen:1998,upa98,nad00,ji01,ray03,par02,Perez:2004,Stokmaier:2008} However, different models~\cite{Soulen:1998,upa98,Strijkers:2001,maz01,Martin-Rodero:2001} describing the transport through S/F interfaces yielded varying values for $P$, also depending on the contact fabrication and the transport regime,~\cite{woo04} an issue that is not yet understood in detail.~\cite{Chalsani:2007} In the following, we want to review the main ideas of two most prominent models shortly and compare the results of both to the same set of experimental data obtained on nanostructured Al/Fe contacts.~\cite{Stokmaier:2008} 

The theoretical analysis of most S/F point-contact experiments has been carried out in the spirit of the Blonder-Tinkham-Klapwijk (BTK) theory \cite{blo82} for Andreev reflection at an interface between N and classical S with spin-singlet pairing. This is the coherent process by which an electron from N enters S and a hole of opposite spin is retro-reflected, creating a spin-singlet Cooper pair in S. Possible ordinary reflection at the S/F interface barrier is parametrised by a phenomenological parameter, the barrier strength $Z$. The sensitivity of the Andreev process to the spin of the carriers originates from the conservation of the spin direction at the interface. Consequently, when there is an imbalance in the number of spin-up and spin-down electrons at the Fermi level, as it is the case in the spin-polarised situation of a ferromagnetic metal, this leads to a reduction of the Andreev reflection probability.~\cite{jon95} Andreev reflection is limited by the minority carriers of the metal. 

In the simpliest approach applied for the analysis of several experiments,~\cite{Soulen:1998,nad00,ji01,ray03,par02, Strijkers:2001} the total current $I$ through the constriction is decomposed  into a fully unpolarised part $(1-P^\prime)I_u$ for which Andreev reflection is allowed and into a fully polarised part $P^\prime I_p$ for which Andreev reflection is zero,
\begin{displaymath}
I=(1-P^\prime)I_u+P^\prime I_p .
\end{displaymath}
The weighting factor $P^\prime$ determines the spin polarisation of the ferromagnet. In the following we will refer to this model as the dispartment model. Consequentially, the conductance $G_{\mathrm{SF}}$ is also decomposed into two parts:
\begin{displaymath}
G_{\mathrm{SF}}=(1-P')G_u(V) +P'G_p(V)
\end{displaymath}
with
\begin{displaymath}
G_{u,p}(V)=\int_{-\infty}^\infty \frac{df(E-V,T)}{dV}[1+A_{u,p}(E,Z)-B_{u,p}(E,Z)]dE
\end{displaymath}
where $G_u$ denotes the conductance, $A_u$ the Andreev reflection probability, and $B_u$ the normal reflection probability of the fully unpolarised channel, and $G_p$, $A_p$, and $B_p$ denote the corresponding quantities of the fully polarised channel. Both contributions are derived in the BTK formalism and following expressions for the zero-temperature conductances $G_u$ and $G_p$ are obtained:~\cite{maz01}
\begin{center}
\begin{tabular}{c|cc}
&$|E|<\Delta$ & $|E|>\Delta$\\
\hline
&&\\
$G_u|_{T=0}(E)$ & $\frac{2(1+\beta^2)}{\beta^2+(1+2Z^2)^2}$ &$\frac{2\beta}{1+\beta+2Z^2}$\\
&&\\
$G_p|_{T=0}(E)$&0&$\frac{4\beta}{(1+\beta)^2+4Z^2}$\\
\end{tabular} 
\end{center}
with $\beta=E/\sqrt{|\Delta^2-E^2|}$.

Despite the attractive simplicity of the BTK formalism it has been shown \cite{xia02,woo04} that application of the BTK formalism (even in its generalized form \cite{maz01}) has certain drawbacks and enforces several assumptions for the analysis. This has mainly to do with the problem to determine $P^\prime$ and $Z$ independently. The physical reason is that both lead to a reduction of the Andreev current and diminish the conductance change in $G=dI/dV$. The model fails to distinguish whether it is high $P$ or high $Z$ that causes the depression of conductance at small bias. This problem is evaded by applying a different theoretical approach.~\cite{Martin-Rodero:2001}

Cuevas and coworkers~\cite{Martin-Rodero:2001, Perez:2004} developed a model based on quasiclassical Green functions. The central quantities of the model are two transmission coefficients $\tau_{\uparrow ,\downarrow}=|t_{\uparrow ,\downarrow}|^2$. Therefore, we will refer to this model as the $\tau_{\uparrow}$-$\tau_{\downarrow}$-model throughout this paper. The transmission coefficients contain all microscopic properties relevant for the transport through the constriction, i.\,e., they account for the majority- and minority-spin bands in the ferromagnet, the electronic structure of the superconductor, and the interface. $t_{\uparrow ,\downarrow}$ and $ r_{\uparrow ,\downarrow}=\sqrt{1-\tau_{\uparrow ,\downarrow}}$, respectively, are the spin-dependent transmission and reflection amplitudes, respectively, entering the normal-state scattering  matrix $\hat S$ which supplements the boundary conditions of the theory. Of course, the restriction to a single conduction channel per spin direction is a rough simplification of the point-contact, but it is finally justified by the agreement with the experiment.~\cite{Perez:2004,Stokmaier:2008} Following the calculation by Cuevas and coworkers the spin-dependent current through the S/F point contact can be separated in two spin contributions, 
\begin{displaymath}
I_{\mathrm{SF}}=I_\uparrow +I_\downarrow
\end{displaymath}
and each contribution can be written in its BTK form~\cite{blo82}
\begin{displaymath}
I_\sigma=\frac{e}{h}\int_{-\infty}^\infty d\epsilon [f(\epsilon-eV)-f(\epsilon)][1+A_\sigma(\epsilon)-B_\sigma(\epsilon)],
\end{displaymath}
where $f(E)$ is the Fermi function, $A_\sigma(\epsilon)$ and $B_\sigma(\epsilon)$ are the spin-dependent Andreev reflection and normal reflection probabilities, respectively, and $\sigma = \uparrow$ or $\downarrow$.  $A_\sigma(\epsilon)$ ($B_\sigma(\epsilon)$) is calculated from the spin-dependent transmission (reflection) amplitudes, and finally, the zero-temperature conductance of the S/F contact adopts the form
\begin{displaymath}
G_{\mathrm{SF}}=\frac{4e^2}{h}\left\{
\begin{array}{ll}
\frac{\tau_\uparrow \tau_\downarrow}{(1+r_\uparrow r_\downarrow)^2-4r_\uparrow r_\downarrow(eV/\Delta)^2}& \quad eV\leq \Delta \\
\frac{\tau_\uparrow \tau_\downarrow+(\tau_\uparrow + \tau_\downarrow-\tau_\uparrow \tau_\downarrow)\sqrt{1-(\Delta/eV)^2}}{[(1-r_\uparrow r_\downarrow)+(1+r_\uparrow r_\downarrow)\sqrt{1-(\Delta/eV)^2}]^2}& \quad eV> \Delta \\
\end{array}
\right.
\end{displaymath}
while the normal-state conductance is given by
\begin{displaymath}
G_N=\frac{e^2}{h}(\tau_\uparrow+\tau_\downarrow)
\end{displaymath}
It is obvious that the Andreev spectra are determined by a set of three free parameters $\tau_\uparrow$, $\tau_\downarrow$, and $\Delta$. The current spin polarisation $P$ in this model is defined by 
\begin{displaymath}
P=\frac{|\tau_\uparrow-\tau_\downarrow|}{\tau_\uparrow+\tau_\downarrow}.
\end{displaymath}
and can be determined from the fit parameters of an experimental Andreev spectrum. We note, that this expression is symmetric with respect to $\tau_\uparrow$ and $\tau_\downarrow$, therefore, one cannot assign a transmission coefficient to the majority or minority charge carriers in the ferromagnet. However, we expect the high transmissive coefficient $\tau_\downarrow$ to correspond to the minority electrons. In the absence of spin polarisation, i.\,e., $P=0$ for a N/S contact with $\tau_\uparrow=\tau_\downarrow$, above formulae reduce to the well-known BTK result~\cite{blo82}.

\begin{figure*}[ht]
	\begin{center}
	\includegraphics[width=0.93\textwidth, clip=]{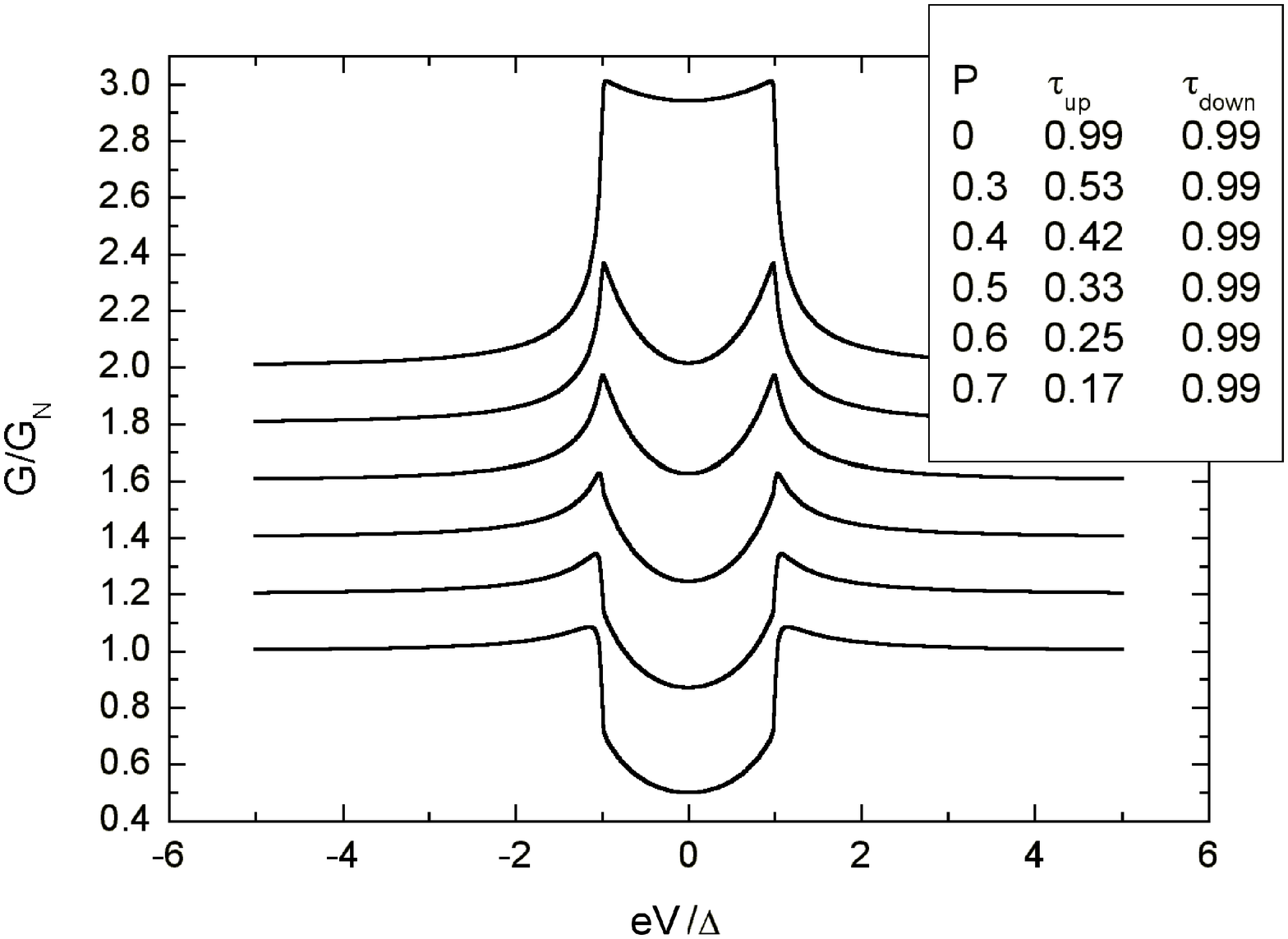}	
\end{center}
	\caption[]{Normalized differential conductance $G/G_{\rm N}$ vs $eV/\Delta$. The curves are calculated with the $\tau_{\uparrow}$-$\tau_{\downarrow}$-model (see text) for different polarisations at $T=0.02 T_c$ . For clarity, the curves are shifted successively upwards by 0.2 units.}
	\label{fig:fig1}
\end{figure*}

Figure \ref{fig:fig1} displays a set of normalized conductance curves for $T=0.02T_c$, $\tau_\downarrow=0.99$, and $\tau_\uparrow=0.99$, 0.53, 0.42, 0.33, 0.25, and 0.17 which equals $P=0$, 0.3, 0.4, 0.5, 0.6, 0.7 from top to bottom. The shape of each spectrum is unambiguously determined by a set of $\tau_\uparrow$, $\tau_\downarrow$. For $P=0$ (top curve) the curve reproduces the well-known BTK result where Andreev reflection causes a doubling of the normal-state conductance for energies $eV<\Delta$. We note that the characteristic double-peak feature at $|eV|=\Delta$ caused by the reduction of the conductance at low bias originates from a small fraction of charge carriers ordinarily reflected at an interface barrier. In the case of two spin-dependent transport channels it is intuitive to consider the high-transmittive spin channel to get a measure for the fraction of charge carriers ordinarily reflected at the interface barrier. From the transmission coefficients  $\tau_{\downarrow}=1/(1+Z_{\downarrow}^2)$ one derives $Z_{\downarrow}=\sqrt{(1-\tau_{\downarrow})/\tau_{\downarrow}}$. For the upper curve which corresponds to the unpolarised BTK case one gets $Z_\downarrow=0.1$ as the parameter equivalent to the BTK interface parameter $Z$.

\begin{figure*}[ht]
	\begin{center}
	\includegraphics[width=0.93\textwidth]{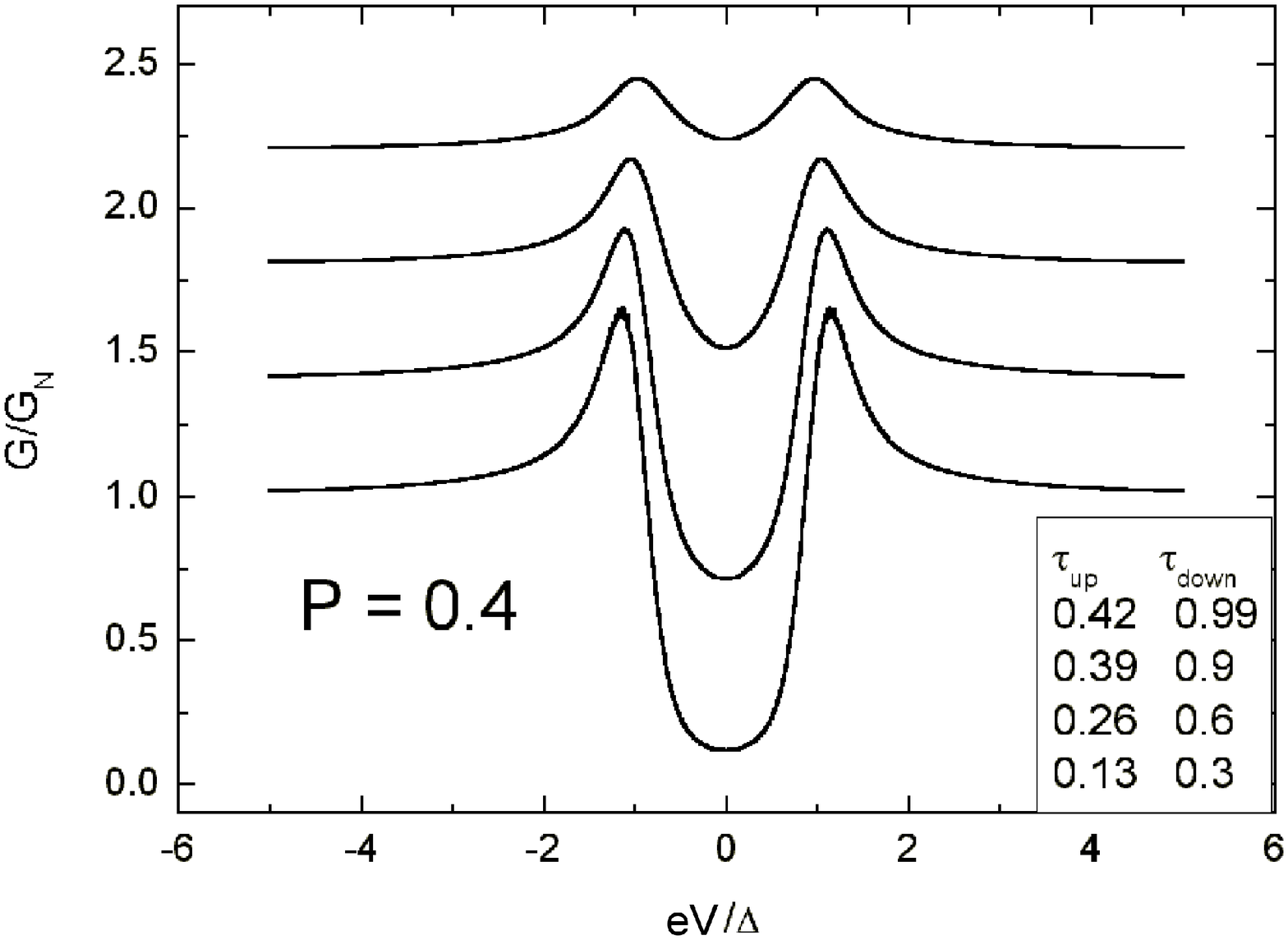}	
\end{center}
	\caption[]{Normalized differential conductance $G/G_{\rm N}$ vs $eV/\Delta$. The curves are calculated with the $\tau_{\uparrow}$-$\tau_{\downarrow}$-model (see text) for the same polarisation $P=0.4$ at $T=0.22 T_c$. For clarity, the curves are shifted successively upwards by 0.4 units. }
	\label{fig:fig2}
\end{figure*}

While the shape of the spectra is unambiguously determined by a set of $\tau_\uparrow$, $\tau_\downarrow$, the spectra for same polarisation can look quite different. Fig.\,\ref{fig:fig2} shows four curves calculated for different sets of $\tau_\uparrow$, $\tau_\downarrow$ which all result in $P=0.4$. The calculations have been performed for a finite temperature $T=0.22 T_c$. The high-transmission spin-channel seems to be decisive whether the curve shape appears more point-contact-like or more tunnelling-like. 

\begin{figure*}[ht]
	\begin{center}
	\includegraphics[width=0.85\textwidth]{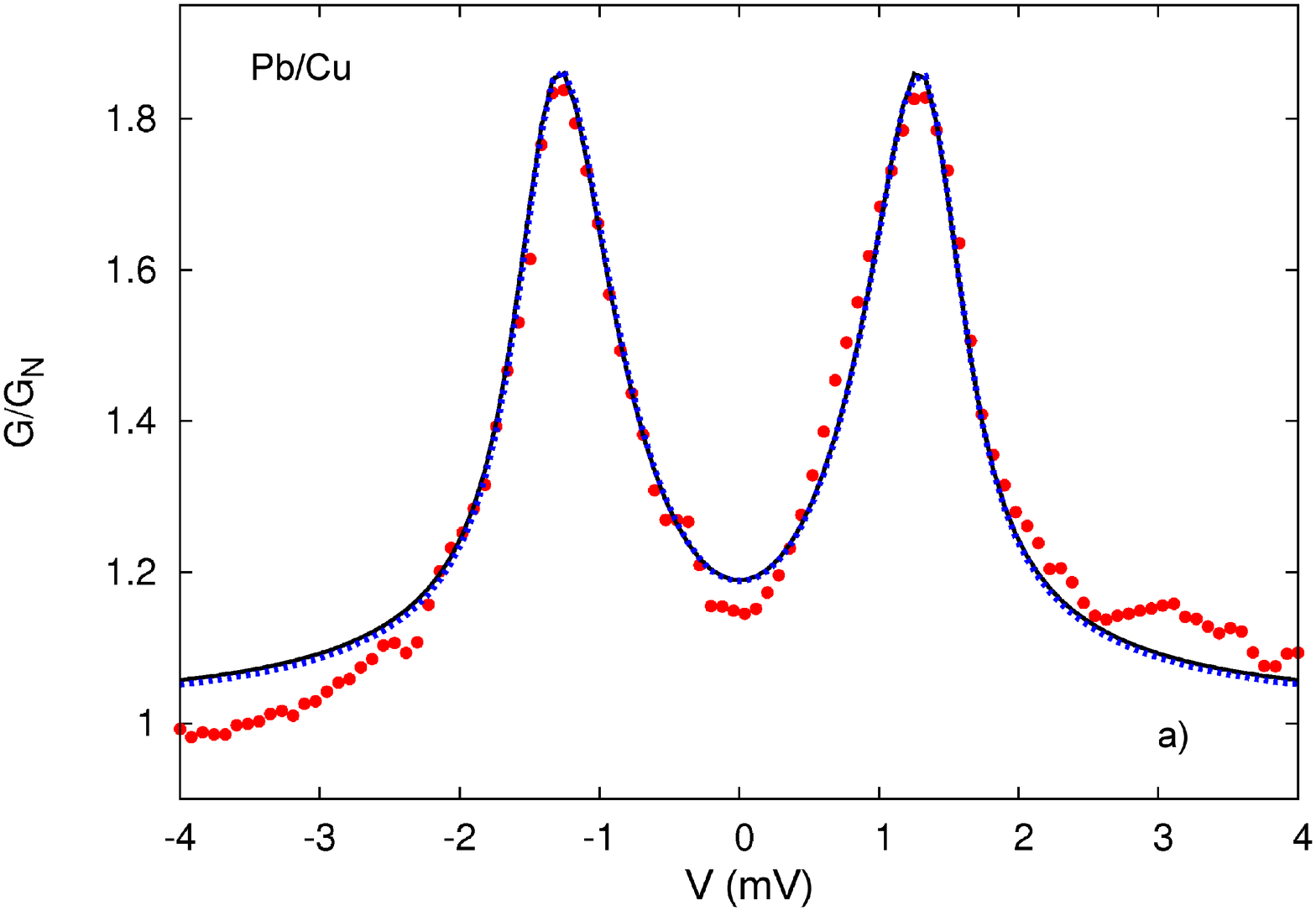}
	\includegraphics[width=0.85\textwidth]{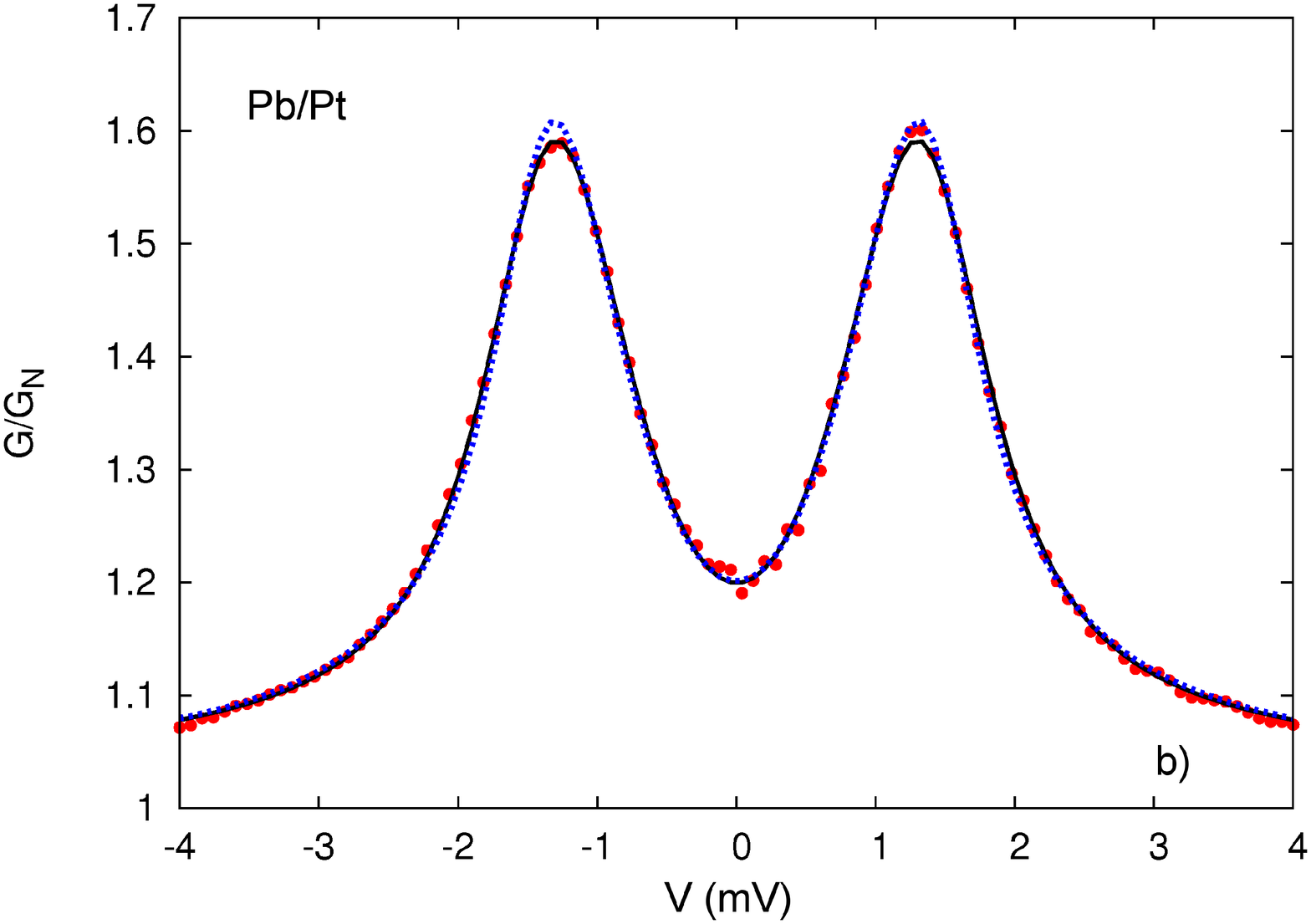}
\end{center}
	\caption[]{Point-contact spectra of S/N contacts  together with fits according to Martin-Rodero \textit{et al.}~\cite{Martin-Rodero:2001} (dashed line) and Mazin  \textit{et al.}~\cite{maz01} (solid line). Upper panel: Pb/Cu contact at 1.4\,K with $R=2.7\,\Omega$. Lower panel: Pb/Pt contact at 2.1\,K with $R=7.5\,\Omega$. For the fit parameters see text.}
	\label{fig:fig3}
\end{figure*}

Before we compare curves calculated by both models, let us first check the validity of the fitting procedure. For this purpose we measured point-contact spectra of S/N point contacts in a $^4$He cryostat down to $T=1.4$\,K and fitted them with both models. The PCs have been established in the edge-to-edge configuration with a sharpened Pb electrode and a normal-metal electrode made from Cu or Pt. Therefore, for both fits we expect $P^\prime=P=0$. Figs.\,\ref{fig:fig3}a and b show the experimental data together with the fits according to the $\tau_{\uparrow}$-$\tau_{\downarrow}$-model~\cite{Martin-Rodero:2001} (dashed line) and the dispartment model~\cite{maz01} (solid line). In both cases we got  almost perfect agreement with the data. For the Pb/Cu contact with $R=2.7\,\Omega$ measured at $T=1.4$\,K (upper panel) we obtained $\Delta_0=\Delta(T=0)=1.38$\,meV, $\tau_{\uparrow}= \tau_{\downarrow}=0.814$ as parameters for the $\tau_{\uparrow}$-$\tau_{\downarrow}$-model, and $\Delta_0=1.38$\,meV, $Z=0.475$, $P^\prime=0.011$ for the dispartment model. For the Pb/Pt contact with $R=7.5\,\Omega$ measured at $T=2.1$\,K (lower panel) the parameters are $\Delta_0=1.48$\,meV, $\Gamma /\Delta_0=0.10$, $\tau_{\uparrow}=0.811$,  $\tau_{\downarrow}=0.819$ for the $\tau_{\uparrow}$-$\tau_{\downarrow}$-model, and $\Delta_0=1.42$\,meV, $\Gamma /\Delta_0=0.10$, $Z=0.488$, $P^\prime=0.015$ for the dispartment model. In both cases we found within the experimental error a good agreement of $P \approx P^\prime \approx 0$ as expected for these non-magnetic metals. The energy gap of Pb determined from the fits coincides fairly good to the gap value reported in literature~\cite{Buckel:2004}. We note that a small broadening of the Pb/Pt spectra caused by inelastic scattering in the contact region is accounted for by introducing the Dynes\cite{dyn84} parameter $\Gamma$ which is of the order of 5-10\% of $\Delta_0$.

\begin{figure*}[ht]
	\begin{center}
	\includegraphics[width=0.93\textwidth,clip=]{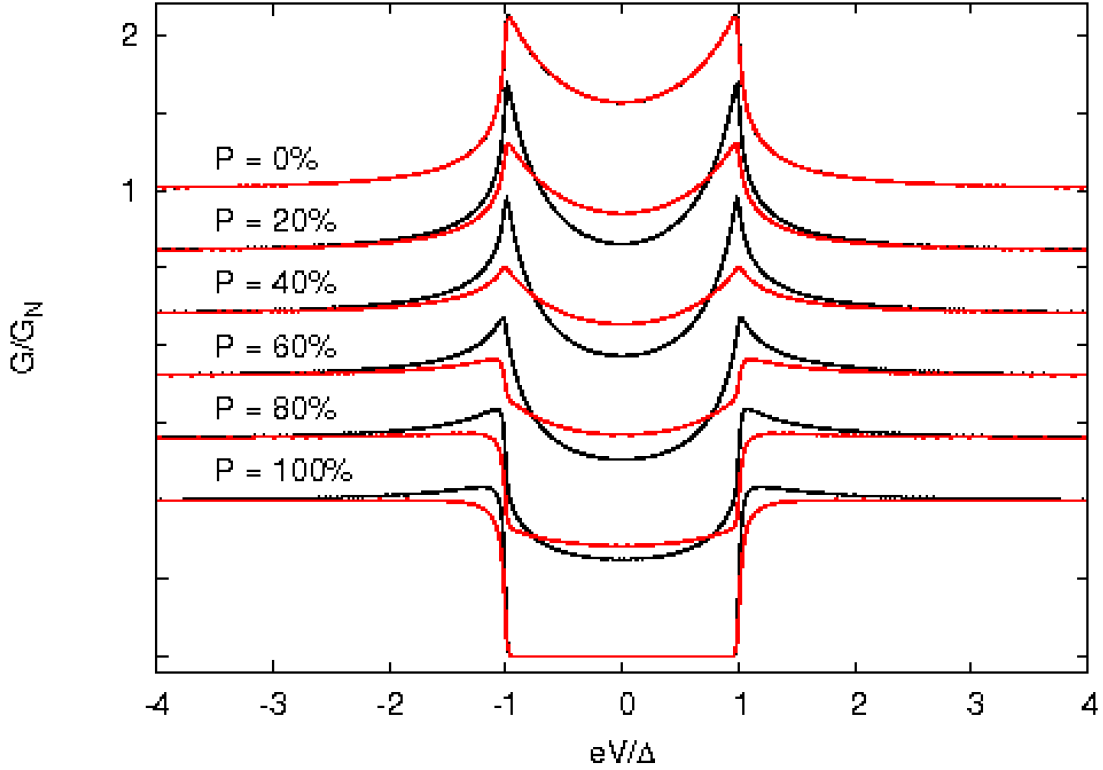}	
\end{center}
	\caption[]{Comparison of Andreev spectra with $P=0$, 0.2, 0.4, 0.6, 0.8, and 1 calculated according to the $\tau_{\uparrow}$-$\tau_{\downarrow}$-model (black lines) and the dispartment model (red lines). For clarity, the curves are shifted successively downwards by 0.4 units.}
	\label{fig:fig4}
\end{figure*}

In the next step we compare curves calculated with the simple dispartment model to those calculated with the $\tau_{\uparrow}$-$\tau_{\downarrow}$-model for nominal same polarisation $P=P^\prime$. Fig.\,\ref{fig:fig4} displays a set of normalized conductance curves calculated for $T=0.1$\,K, $P=P^\prime=0$, 0.2, 0.4, 0.6, 0.8, and 1, and $Z=0.3$ for the  dispartment model, and $\tau_\downarrow=0.917$ for the $\tau_{\uparrow}$-$\tau_{\downarrow}$-model, respectively, which corresponds to $Z_{\downarrow}=\sqrt{(1-\tau_{\downarrow})/\tau_{\downarrow}}=0.3$. In the extreme case $P=P^\prime =0$ which describes a S/N contact the calculations perfectly agree with each other, as well as for the other extreme case $P=P^\prime =1$ which describes a halfmetallic F/S contact. For the latter, there is only a small difference in the vicinity of the coherence peaks at $|eV/\Delta|=1$. However, at intermediate values with increasing polarisation the Andreev signal is much faster suppressed for the dispartment model than for the $\tau_{\uparrow}$-$\tau_{\downarrow}$-model.  Our comparison discloses a notable difference of both quantities which makes questionable a contrasting juxtaposition of $P$ and $P^\prime$ values derived from the analysis of experimental data by one of these models.

\begin{figure*}[ht]
	\begin{center}
	\includegraphics[width=0.93\textwidth]{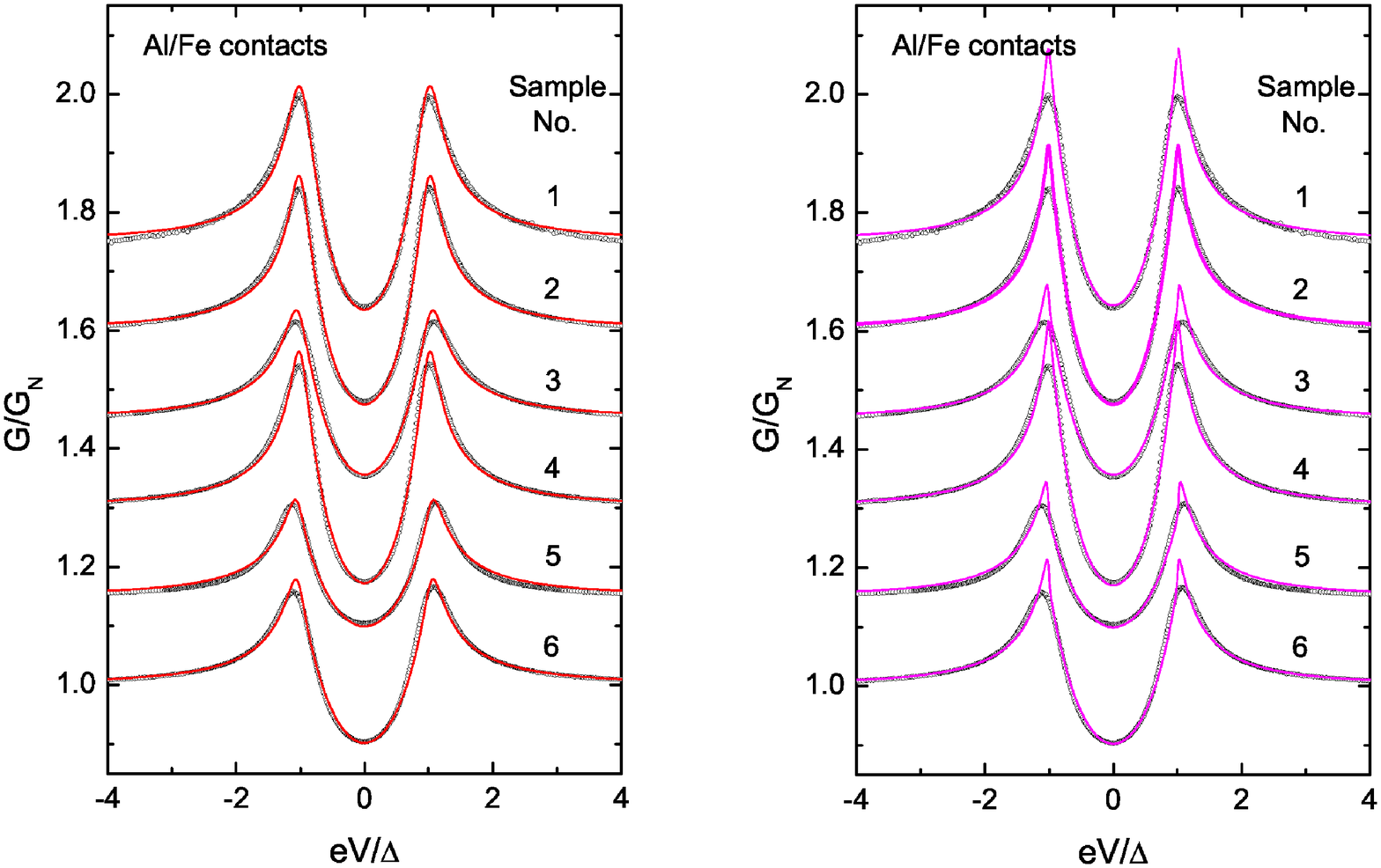}
\end{center}
	\caption[]{Point-contact spectra of Al/Fe contacts at 0.1\,K together with fits according to the $\tau_{\uparrow}$- $\tau_{\downarrow}$-model (left panel) and according to the dispartment model (right panel).}
	\label{fig:fig5}
\end{figure*}

In order to illustrate this difference on experimental data we used both models to fit the same set of data measured on Al/Fe nanostructured point contacts.~\cite{Stokmaier:2008} The PCs were fabricated by structuring a hole of 5 - 10\,nm diameter with electron-beam lithography and subsequent reactive ion etching into a 50-nm thick Si$_3$N$_4$ membrane, evaporating on one side a 200\,nm Al layer, and on the other side a 12-nm thick Fe layer and a Cu layer of thickness $d_{\rm Cu} = 188$\,nm as a low-ohmic electrode.~\cite{nano} Fig.\,\ref{fig:fig5} displays the normalized conductance spectra of six different nanostructured PCs with contact resistances between $2.7\,\Omega$ and $24.2\,\Omega$ measured in a dilution refrigerator at $T=0.1$\,K together with fitting curves calculated with the $\tau_{\uparrow}$-$\tau_{\downarrow}$-model (left panel) and the dispartment model (right panel). Both models perfectly describe the data, minor deviations are observed only at $|eV/\Delta|\approx 1$ where the experimental curves are more rounded probably caused by a leveling-off of the electron temperature due to heating by electromagnetic stray fields or a small pair-breaking effect by Fe. The corresponding fit parameters are listed in Table\,\ref{table:parameters}. 
\begin{table*}[t]
\begin{tabular}{cc|ccccc|ccc}
Sample& $R_N$ & $\Delta_0$ &$\tau_{\uparrow}$&$\tau_{\downarrow}$& $Z_{\downarrow}$ & $P$ & $\Delta_0$ & $Z$ & $P^\prime$\\
No.&$(\Omega)$&(meV)&&&&&(meV)&&\\
\hline
1 & 2.68 & 0.174 & 0.371 & 0.983 &0.132 & 0.452 & 0.176 & 0.337 & 0.396 \\
2 & 6.98 & 0.175 & 0.362 & 0.984 &0.128 & 0.460 & 0.176 & 0.338 & 0.407 \\
3 & 7.29 & 0.157 & 0.349 & 0.993 &0.083 & 0.480 & 0.158 & 0.272 & 0.444 \\
4 & 9.59 & 0.190 & 0.361 & 0.984 &0.128 & 0.463 & 0.191 & 0.342 & 0.407 \\
5 & 18.4 & 0.166 & 0.348 & 0.997 &0.054 & 0.482 & 0.168 & 0.218 & 0.497 \\
6 & 24.2 & 0.174 & 0.343 & 0.994 &0.078 & 0.487 & 0.174 & 0.262 & 0.494 \\
\end{tabular}
\caption[]{Fit parameters $\Delta_0$, $\tau_{\uparrow}$, and $\tau_{\downarrow}$ of the $\tau_{\uparrow}$-$\tau_{\downarrow}$-model and  $\Delta_0$, $Z$, and $P^\prime$ of the dispartment model for conductance spectra of nanostructured Al/Fe point contacts. From the transmission coefficients $\tau_{\uparrow}$ and $\tau_{\downarrow}$ the interface barrier $Z_{\downarrow}$ and the current spin polarisation $P$ have been calculated.}
\label{table:parameters}
\end{table*}
Within an uncertainty of 1\% the same gap value $\Delta_0$ is found for both models, however, there is a notable difference in the $Z$ parameter, which is a factor 2-3 higher in the dispartment model, and the spin polarisation $P^\prime$ which is lower. Although the origin of $Z$ is not clear at all,  in both models it subsumes all ordinary reflection of charge carriers that occurs at the interface for $P=P^\prime=0$, e.g., reflection caused by an insulating interface layer, lattice imperfections, Fermi velocity mismatch, etc.. For $P$ and $P^\prime>0$ the situation is less apparent. The physical reason is that $Z$ and $P^\prime$ both lead to a reduction of the Andreev current. The dispartment model obviously fails to distinguish whether a high spin polarisation or a high barrier causes the depression whereas for the $\tau_{\uparrow}$-$\tau_{\downarrow}$-model as per definition only the conductance channel not affected by the suppression of Andreev reflection due to polarisation is considered to determine ordinary reflection.

Another important aspect is that the previously reported dependence of the spin polarisation on the contact size~\cite{Stokmaier:2008} is robust against the model used to extract $P$. Independently of the model there is a clear reduction of the spin polarisation with decreasing contact resistance $R_N$, i.\,e. increasing contact radius $a$. The reduction of $P$ has been allocated as being due to spin-orbit scattering in the contact region with a constant scattering length $\ell_{so}$ modelled by a simple exponential decay~\cite{Stokmaier:2008}, $P(a) = P_0\exp{(-a/\ell_{so})}$. A spin-orbit scattering length $\ell_{so}  = 255$\,nm has been obtained for the analysis with the $\tau_{\uparrow}$-$\tau_{\downarrow}$-model. The same systematic trend of $P^\prime(a)$  is found for the dispartment model albeit with a lower value for $\ell_{so}$. For small contacts both models result in almost the same spin polarisation.

In summary, we have discussed possible reasons for the scatter of polarisation values found for the spin polarisation measured by Andreev reflection in point-contact experiments. We showed that the scatter is partially caused by the models used to extract the spin polarisation from the data, and partially caused by intrinsic mechanisms like the spin-orbit scattering in the contact region.

\begin{acknowledgments}
The authors thank H. v. L\"ohneysen for stimulating discussions and continuous support of the point-contact research at the Physikalisches Institut. G.G. acknowledges the fruitful and long-lasting collaboration over almost two decades with Igor Yanson who brought him into touch with point-contact spectroscopy during a sabbatical stay at Karlsruhe. Igor Yanson was a exceptionally gifted teacher and outstanding scientist. We acknowledge the financial support provided within the DFG-Center for Functional Nanostructures.
\end{acknowledgments}

\end{document}